# Markov-Binary Visibility Graph: a new method for analyzing Complex Systems

Y. Sadra *, Z. Arasteh Fard, S. Ahadpour

*Department of Sciences, University of Mohaghegh Ardabili, Ardabil, Iran.*
*Email address:* yasersadra@gmail.com

**Abstract**

In this work, we introduce a new and simple transformation from time series to complex networks based on markov-binary visibility graph(MBVG). Due to the simple structure of this transformation in comparison with other transformations be obtained more precise results. Moreover, several topological aspects of the constructed graph, such as degree distribution, clustering coefficient, and mean visibility length have been thoroughly investigated. Numerical simulations confirm the reliability of markov-binary visibility graph for time series analysis. This algorithm have the capability of distinguishing between uncorrelated and correlated systems. Finaly, we illustrate this algorithm analyzing the human heartbeat dynamics. The results indicate that the human heartbeat (RR-interval) time series of normally, Congestive Heart Failure (CHF) and Atrial Fibrillation (AF) subjects are uncorrelated, chaotic and correlated stochastic systems, respectively.

The visibility algorithm has more applications in various fields such as stock market indices[13–15], human strive intervals[16], human heartbeat dynamics[17,18], energy dissipation rates in three-dimensional fully developed turbulence[19], foreign exchange rates[20], binary sequences[21]. The visibility algorithm transforms a time series into a visibility graph[5]. Lacasa et al. used geometrical visibility of top of the bars in the time series's bar diagram to produce visibility graphs of time series. On the other hand, there is a linear interpolation between data which have mutual visibility[25]. This algorithm depends on the statistical persistence of the time series[25]. Therefore, this Dependence is less, if visibility be low. one of these methods is horizontal visibility graph[24]. In this way, they used of limit visibility range from $(0, 2\pi)$ into $\{0, \pi\}$ are able to good discriminate uncorrelated randomness from chaotic and correlated randomness series[24,26]. Another methods, used of fluctuations of the time series's line diagram for time series analysis. on the other words, we transform fluctuations of the time series's line diagram into two-state markov chain and then using binary visibility graph, we map the two-state markov chain into complex networks. We call this algorithm as markov-binary visibility algorithm(MBVA). This algorithm is less dependent on the statistical persistence of the time series in comparison with horizontal visibility algorithm. Therefore, this algorithm be obtained more precise results. The work is organized as follows. In Sec.II we introduce the MBVG algorithm. In Sec.III we derive results for statistical properties of the MBVG such as degree distribution, clustering coefficient, visibility length and mean visibility and finally, in Sec.IIII, we illustrate this algorithm analyzing the human heartbeat dynamics. Finally, we conclude the work.

## 2  Structure of Markov-Binary Visibility Graph

A Markov chain is a discrete-time process for which the future behavior, given the past and the present, only depends on the present and not on the past. Also, a Markov chain is characterized by a set of states $S$ and the transition probabilities $P_{i \to j}$ between the states. Using this properties, we show how Markov-Binary Visibility Algorithm (MBVA) maps the time series into a complex network (see Fig. 1). Here, we briefly describe the markov-binary visibility algorithm:

First, to consider $\{y_i\}_{i=1,...,N}$ be a time series of N data. Next, to draw line diagram corresponding to time series. Then, we get slope $(m)$ between any two consecutive data in the time series's line diagram (see Fig. 1). Now, the set of states $S$ of markov chain is defined as follows:

$$S = \begin{cases} up & m > 0 \\ down & m \leq 0 \end{cases} \quad (1)$$

Then, to use of states $S$ of two-state markov chain obtained fluctuations of line diagram corresponding to time series, we construct two-state markov chain of states $up$ and $down$. We replace states $up$ and $down$ of this chain with 1 and 0 in binary sequence, respectively. Thus, we have Markov-Binary Sequence of $N$ bits($MBS = \{x_i\}_{i=1,...,N}$). Since in chaotic time series, the number of $up$ states is always more than number of $down$ states, thereby, because complexity of MBVG corresponding to chaotic time series be Maintained, the set of states $S$ of markov chain is defined as follows:

$$S = \begin{cases} up & m \geq 0 \\ down & m < 0 \end{cases}$$

Then, we replace states $up$ and $down$ of this chain with 0 and 1 in markov-binary sequence, respectively. With regard to the above contents, The markov-binary visibility algorithm assigns each fluctuations of line diagram corresponding to time series to a bit of the markov-binary sequence and then assigns each bit of the markov-binary sequence to a node in the MBVG. Two nodes $i$ and $j$ in the MBVG are connected if one can draw a visibility line in the binary sequence joining the neighboring $x_i$ and $x_j$ that does not intersect any intermediate bits height. $x_i(x_j)$ can only be 0 or 1. Therefore, $i$ and $j$ are two connected nodes if the succeeding geometrical criterion is satisfied within the binary sequence:

$$x_i + x_j > x_n \quad that \quad x_n = 0 \quad for \quad all \quad n \quad such \quad that \quad i < n < j \ . \tag{2}$$

It is important to note that, MBVA is a new version of its associated visibility algorithm; yet, it is both geometrically simpler and analytically more solvable than the visibility algorithm introduced by [5]. The MBVG associated with a Time series is always connected and undirected. In what follows, we will show the geometrical simplicity of the markov-binary version of the visibility algorithm and that it grants an analytically easier solubility; this new method can attest to perfect distinguish between uncorrelated systems and correlated systems(e.g. chaotic and correlated stochastic). Also, the MBVG is able to analysis of the fractal time series that will be explored in detail in further work. This point should also be noted that a MBVG, like a visibility graph is invariant under affine transformation of the series data since the visibility criterion is invariant under re-scaling of both horizontal and vertical axes, and under horizontal and vertical transformation [27]. MBVG is a planar graph which is connective, undirected and a combination of two linear and maximal planar subgraphs(see Fig. 1).
The constituting subgraphs are:
1. Maximal Planar Graph($MPG_n$): It is a connective simple planar graph. This graph is not a simple planar graph on adding an extra edge $e$ (i.e., $e \notin E(MPG)$ ). $n$ and $E(MPG)$ are the number of nodes and edges of the graph, respectively.

2. Linear Planar Graph($LPG_n$): It is a connective planar graph without loops and multiple edges. The degree of all the nodes is 2 except the first and the last nodes which have a degree of 1.

As will be discussed later, MBVA can be employed as a new strong analyst for time series analysis. In order to show the significance of MBVG, the analysis of its statistical properties is necessary.

## 3 Statistical properties of the MBVG corresponding with a time series

Markov-binary visibility graphs inherit the statistical(topological) properties of the associated time series. Analyzing MBVG from statistical point of view, we studied its visibility degree and length probability distribution, clustering coefficient, and mean visibility length. As will be shown in this work, these properties will be proven to have the useful capability for analysis of time series.

### 3.1 Probability distribution for the visibility degree P(k)

As we mentioned before, a markov chain is characterized by a set of states $S = \{up, down\}$ and the transition probabilities $P_{i \to j}$ between the states. consequently, a markov-binary sequence of time series is characterized by a set of states $S = \{0, 1\}$ and the transition probabilities $P_{i \to j}$ between the states. Here, $P_{i \to j}$ is the probability that the markov-binary sequence is at the next time point in state $j$, given that it is at the present time point at state $i$. The matrix $P_{MBS}$ with elements $P_{i \to j}$ is called the transition probability matrix of the markov-binary sequence and as be defined follow[29]:

$$P_{MBS} = \begin{pmatrix} P_{1 \to 1} & P_{1 \to 0} \\ P_{0 \to 1} & P_{0 \to 0} \end{pmatrix} = \begin{pmatrix} (1-p) & p \\ q & (1-q) \end{pmatrix}. \tag{3}$$

Let us consider a markov-binary sequence of time series such that $MBS \in \{0, 1\}$. Considering the bits of $MBS = \{x_i\}_{i=1,...,N}$, we take an arbitrary bit according to which, we are going to define the degree distribution. This bit is the one to which, all the other bits are connected by "visibility line". Therefore, we call that the "observer" bit here after with the label $x_0$. In order to obtain the degree distribution $P(k)$ [28] of the associated graph, we are going to estimate the probability if $x_0$ can observe $k$ other bits. $k$ is the degree of the MBVG. For any configuration of the degree $k$, there always exist two bounding bits both with the value 1. This implies that the minimum possible visibility degree is $k = 2$. We call the bits between the bounding bits and the observer as the inner bits: some can be observed by the observer called the "visible bits" and some can not be observed called the "nonvisible bits". It is worthwhile to mention that although both the visible and nonvisible bits are counted in calculating the degree distribution probability, it is only the visible bits who determine the degree of the visibility graph. In line with what is mentioned above, for a given $k$, there are exactly $k - 1$ different possible configurations $\{F_i\}_{i=0,\ldots,k-2}$, where the index $i$ determines the number of visible bits on the left-hand side of $x_0$ (see Fig. 2). According to the nature of the configurations, the case where $k = 4$ and $x_0 = 0$ has a different behavior, since the observer should always remain between two visible bits. Also, for $k \geq 5$, the observer can only be the bit with value 1. It is due to the fact that the observer $x_0 = 0$ may not observe more than 4 bits.

In order to make the construction of the $F_i$ more clear, we have calculated as an example, a set of possible configurations for $k = 4$ with the results denoted in Fig. 2. In $x_0 = 1$, $F_0$ is the configuration where, none of the $k - 2 = 2$ visible bits are located on the left of $x_0$, hence the left bounding bit is labeled as $x_{-1}$ and the right bounding bit is labeled as $x_3$. For $x_0 = 0$, $F_0$ is the configuration where one of the $k - 2 = 2$ visible bits is located on the left of $x_0$, so the left bounding bit is $x_{-2}$ and the right bounding bit accounting for $n$ nonvisible bits in the way, is labeled as $x_{n+1}$. In $x_0 = 1$, $F_1$ is the configuration for which, $x_0$ is located between two visible bits. For $x_0 = 0$, $F_1$ is the configuration for which, $n_1$ nonvisible bits are located on the left of $x_0$ and $n_2$ nonvisible bits are located on the right. Finally, in $x_0 = 1$, $F_2$ is the configuration for which, both visible bits are located on the left of the observer. For $x_0 = 0$, $F_2$ is the configuration where one of the $k - 2 = 2$ visible bits is located on the right of $x_0$, and therefore the right bounding bit is labeled as $x_2$ and the left bounding bit is labeled as $x_{-(n+1)}$. Notice that $n$ nonvisible bits can be located on the right of the observer bit (see Fig. 2).

Consequently, $F_i$ corresponds to the configuration for which $i$ visible bits are placed on the left of $x_0$, and $k - 2 - i$ visible bits are placed on the right. Each of these possible configurations have an associated probability $P_{F_i}(x_0)$ that will result in $P(k)$ such that

$$P(k) = \sum_{i=0}^{k-2} P_{F_i}(x_0). \tag{4}$$

For $k \geq 2$, the total probability that $x_0$ observes is 1:

$$\sum_{k \geq 2} P(k) = 1$$

To consider in Eq. 4, $q = p_1$ and $p = (1 - p_1)$, thus, we have,

$$P_{MBS} = \begin{pmatrix} p_1 & (1 - p_1) \\ p_1 & (1 - p_1) \end{pmatrix} \qquad (5)$$

where $p_1$ and $(1 - p_1)$ are probability of bits of 1 and probability of bits of 0 in the markov-binary sequence, respectively. Note that the definition of the $P_{i \to j}$ implies that the row sums of $P_{MBS}$ are equal to 1 [29].
We show that the general relation for $P(k)$ as following:

$$P(k) = g(k) \times T_{P_{MBS}}(k) \qquad (6)$$

where $g(k)$ and $T_{P_{MBS}}(k)$ are the probability density function and the total of possible configurations probability from degree of $k$ (that is function from multiplying elements of transition probability matrix $P_{MBS}$), respectively. Here, the probability density function $g(k)$ is equal to 1 ($g(k) = 1$) unless value of degree $k$ is more than or equal to 5 i.e. $k \geq 5$, In which case the $g(k)$ can de defined as follows:

$$g(k \geq 5) = \begin{cases} 1 & \text{uncorrelation systems} \\ \alpha \, e^{-(\sigma^2(k-2))^d} & \text{correlation systems (stochastic and chaotic systems)} \end{cases} \qquad (7)$$

where $\alpha$ and $\sigma^2$ are constant value and variance of the markov-binary sequence, respectively. The value of $d$ is proportional with correlation exponent $\tau$ in correlation stochastic systems and correlation dimension $D$ in chaotic systems. Starting with $k = 2$, i.e. the probability that the observer bit has two and only two visible bits:
For $x_0 = 0$:

$$P_{F_0}(x_0 = 0) = P(1 \underbrace{0}_{x_0} 1) = P(1) \times P_{1 \to 0} \times P_{0 \to 1} = p_1^2 (1 - p_1)$$

For $x_0 = 1$:

$$P_{F_0}(x_0 = 1) = P(1 \underbrace{1}_{x_0} 1) = P(1) \times P_{1 \to 1} \times P_{1 \to 1} = p_1^3$$

Then,

$$T_{P_{MBS}}(2) = P_{F_0}(x_0 = 0) + P_{F_0}(x_0 = 1)$$

and

$$P(k = 2) = g(2) \times T_{P_{MBS}}(2) = p_1^2 \qquad (8)$$

The second case is $k = 3$, i.e., for the observer which can see three and only three visible bits. In this way, we encounter two different configurations: $F_0$ and $F_1$. As can be seen from Fig. 2, for $x_0 = 0$, an arbitrary number of nonvisible bits($n$ for $F_0$ and $n'$ for $F_1$) are located between the observer and the bounding bit. It is crucial to note that the nonvisible bits must be taken into account in the probability calculation. For $F_0$, the nonvisible bits are $b_j = 0 (j=1,\cdots,n)$ and for $F_1$, they are $d_j = 0 (j=1,\cdots,n')$. Hence,

$$P_{F_0}(x_0 = 0) = P(1\underbrace{0}_{x_0}\overbrace{00...00}^{b_1,...,b_n}1) = P(1) \times P_{1\to 0} \times P_{0\to 0} \times \overbrace{P_{0\to 0} \times ... \times P_{0\to 0}}^{n} \times P_{0\to 1}$$

$$P_{F_1}(x_0 = 0) = P(1\overbrace{00...00}^{d_1,...,d_n}0\underbrace{0}_{x_0}1) = P(1) \times P_{1\to 0} \times \overbrace{P_{0\to 0} \times ... \times P_{0\to 0}}^{n} \times P_{0\to 0} \times P_{0\to 1},$$

Considering all the configurations for $x_0 = 0$ with and without nonvisible bits($F_0$ without nonvisible bits, $F_0$ with a single nonvisible bit, $F_0$ with two nonvisible bits, and so on, and the same for $F_1$):

$$P_{F_0}(x_0 = 0) = p_1^2(1-p_1)^2[1 + \sum_{n=1}^{\infty}(1-p_1)^n] = p_1(1-p_1)^2$$

Where the first term in the square bracket in Eq.(4) corresponds to the contribution of a configuration with no nonvisible bits and the second is a sum over the contributions of $n$ nonvisible bits. For $x_0 = 1$, there are no nonvisible bits therefore:

$$P_{F_0}(x_0 = 1) = P(1\underbrace{1}_{x_0}01) = P(1) \times P_{1\to 1} \times P_{1\to 0} \times P_{0\to 1} = p_1^3(1-p_1)$$

Similar results can be reached for $P_{F_1}$, then

$$T_{P_{MBS}}(k=3) = 2(P_{F_0}(x_0=0) + P_{F_0}(x_0=1)).$$

Consequently, one gets

$$P(k=3) = g(3) \times T_{P_{MBS}}(3) = 2p_1(1-p_1)(p_1^2 - p_1 + 1) \tag{9}$$

The third step is to calculate the probability for $k = 4$, i.e., for the observer which has four and only four visible bits. There are three different configurations: $F_0$, $F_1$, $F_2$. This case is different from the others in that, for $x_0 = 0$, there always exists a visible bit between the observer and the boundary. This stems from the nature of observer and the geometrical criterion for $k = 4$. Accordingly, for $F_0[F_2]$, all the nonvisible bits $b_j = 0(j=2,\cdots,n)[b_i = 0(i=-2,\cdots,-n)]$ can be on the right[left] of the observer. With respect to the symmetry in the visibility geometrical criterion of BVG, it will be sufficient to calculate the probability for one configuration:

$$P_{F_0}(x_0 = 0) = P(10\underbrace{0}_{x_0}0\overbrace{00...00}^{b_1,...,b_n}1)$$

$$= P(1) \times P_{1\to 0} \times P_{0\to 0} \times P_{0\to 0} \times \overbrace{P_{0\to 0} \times ... \times P_{0\to 0}}^{n} \times P_{0\to 1},$$

From which, we can get:

$$P_{F_0}(x_0 = 0) = p_1^2(1-p_1)^3[\sum_{n=1}^{\infty}(1-p_1)^n] = p_1(1-p_1)^4$$

However, for $F_1$, $n_1$ nonvisible bits $b_i = 0$(i=-2,$\cdots$,$-n_1$) are to the left of the observer and $n_2$ nonvisible bits $b_j = 0$(j=2,$\cdots$,$n_2$) are to the right. Hence,

$$P_{F_1}(x_0 = 0) = P(1\overbrace{0...0}^{b_1,...,b_{n_1}}0\underbrace{0}_{x_0}0\overbrace{00...00}^{b_1,...,b_{n_2}}1)$$

$$= P(1) \times P_{1\to 0} \times \overbrace{P_{0\to 0} \times ... \times P_{0\to 0}}^{n_1} \times P_{0\to 0} \times P_{0\to 0} \times \overbrace{P_{0\to 0} \times ... \times P_{0\to 0}}^{n_2} \times P_{0\to 1},$$

From which, we can get:

$$P_{F_1}(x_0 = 0) = p_1^2(1-p_1)^3[1+\frac{1}{2}\sum_{n_1=1}^{\infty}(1-p_1)^{n_1}+\frac{1}{2}\sum_{n_2=1}^{\infty}(1-p_1)^{n_2}] = p_1(1-p_1)^3$$

The first term corresponds to the state without nonvisible bits while the next two terms represent the configurations with nonvisible bits. Regarding the symmetry in the BVG visibility geometrical criterion for $x_0 = 1$, we may write $P_{F_0} = P_{F_1} = P_{F_2}$. Therefore, it suffices to calculate $P_{F_0}(x_0 = 1)$:

$$P_{F_0}(x_0 = 1) = P(1\underbrace{1}_{x_0}001) = P(1) \times P_{1\to 1} \times P_{1\to 0} \times P_{0\to 0} \times P_{0\to 1}$$

$$= p_1^3(1-p_1)^2$$

and then,

$$T_{P_{MBS}}(k = 4) = 3P_{F_0}(x_0 = 1) + 2P_{F_0}(x_0 = 0) + P_{F_1}(x_0 = 0).$$

Hence, it is straightforward to obtain:

$$P(k = 4) = g(4) \times T_{P_{MBS}}(4) = 3p_1(1-p_1)^2(\frac{5}{3}p_1^2 - \frac{5}{3}p_1 + 1) \qquad (10)$$

The next case at issue is $k \geq 5$. Inasmuch as the zero observer($x_0 = 0$) may not see more than four bits in BVG, we are not going to witness $x_0 = 0$ when $k \geq 5$. Consequently, in all the configurations for $k \geq 5$, we are only left with the observer $x_0 = 1$.

When $k = 5$, we can only have three inner bits so, there will be four configurations(see Fig.2).

$$P_{F_0}(x_0 = 1) = P(1\underbrace{1}_{x_0}0001) = P(1) \times P_{1 \to 1} \times P_{1 \to 0} \times P_{0 \to 0} \times P_{0 \to 0} \times P_{0 \to 1}$$

$$= p_1^3(1 - p_1)^3$$

and

$$T_{P_{MBS}}(k = 5) = 4P_{F_0}(x_0 = 1)$$

Ultimately, one gets

$$P(k = 5) = g(5) \times T_{P_{MBS}}(k = 5) = 4g(5)\, p_1^3(1 - p_1)^3 \qquad (11)$$

Eventually, the probability distribution for the visibility degree is as follows:

$$P(k) = (k - 1)g(k)\, p_1(1 - p_1)^{k-2} F(p_1) \qquad (12)$$

where

$$F(p_1) = \begin{cases} p_1 & k = 2 \\ p_1^2 - p_1 + 1 & k = 3 \\ \frac{5}{3}p_1^2 - \frac{5}{3}p_1 + 1 & k = 4 \\ p_1^2 & k \geq 5 \end{cases}$$

In order to distinguish perfect between systems, we define entropy of degree distribution $P(k)$ as following:

$$H(P(k)) = -\sum_{k=2}^{m} P(k) \ln P(k). \qquad (13)$$

If diagram of the entropy plotted according to the probability of *up* states $P(up)$, values of entropies of the correlated stochastic systems ($H_s$) and uncorrelated systems ($H_{un}$) and chaotic systems ($H_c$) are as follows:

$$H_s < H_{un} < H_c .$$

we check the reliability of the above equations using the following examples:

**Chaotic logistic map :** It is a nonlinear chaotic map as following:

$$y_{n+1} = 4y_n(1 - y_n)$$

which the correlation dimension $D$ is 1.016 [30].

**Henon map :** It is a two-dimension nonlinear chaotic map as following:

$$x_{n+1} = y_n + 1 - ax_n^2,$$

$$y_{n+1} = bx_n,$$

which the correlation dimension $D$ is 1.220 (In this case $a = 1.4$ and $b = 0.3$)[30].

**lozi map :** It is a piecewise linear variant of the Henon map given by

$$x_{n+1} = y_n + 1 - a|x_n|,$$

$$y_{n+1} = bx_n,$$

which the correlation dimension $D$ is 1.384 (In this case $a = 1.7$ and $b = 0.5$)[30].

**Linear piecewise map :** It is a piecewise linear map with uncorrelated specificity which defined as following[31]:

$$x_{n+1} = \begin{cases} \frac{x_n}{p} & 0 \leq x_n < p \\ \frac{x_n - p}{1-p} & p \leq x_n \leq 1 \end{cases}$$

where $p \in [0, 1]$ and the value of $p_1$ in the markov-binary sequence is equal to $p$. ( Here, $p = 0.30$, $p = 0.51$, $p = 0.70$)

**Ornstein-Uhlenbeck (O-U) process (Red noise) :** The O-U process is stationary, Gaussian, and Markovian[32]. It is a Fokker-Planck equation (FPE) as following:

$$dx_t = -\kappa \, x_t \, dt + \sqrt{D} \, dW_t$$

with the damping parameter (correlated exponent) $\kappa = -\frac{1}{\tau}$ ($\tau$ is the "memory" of the system) and diffusion coefficient $D$ and initial condition $p(x, t_0|x_0, t_0) = \delta(x - x0)$ that generate short-range correlated series ($W_t$ denotes the Wiener process)[33] ( Here, $\tau = 0.5$, $\tau = 1.0$, $\tau = 1.5$).

**Long-Range stochastic (L-R) process :** It is a random process with a flat power spectral density that the power-law correlation function is proportional with power of time i.e. $R(t) \sim t^{-\tau}$ [33]. ( Here, $\tau = 0.5$, $\tau = 1.0$, $\tau = 2.0$).

Numerical results shows that the degree distribution $P(k)$ and entropy of degree distribution $H(P(k))$ of the MBVG corresponding to time series of above examples can be predicted by Eqs. 12 and 13 (see Figs. 3, 4, 5). Also, using entropy of degree distribution $H(P(k))$, this algorithm can distinguish uncorrelated systems from correlated systems that including correlated stochastic systems and chaotic systems(see Fig. 5).

## 3.2 Probability distribution of the clustering coefficient P(C)

If the node (i) in a MBVG has $k_i$ edges, it is connective with $k_i$ other nodes. If the first neighbors of this node(i) is a part of a cluster, there are $\binom{k_i}{2}$ edges between the neighbors. The ratio of the number of edges ($E_i$) existing between $k_i$ nodes to all the edges $\binom{k_i}{2}$ gives the clustering coefficient for that node[8,34]. In other words, it is the number of the nodes that observe (i) nodes and have mutual visibility(the triangles) and is normalized by the possible set of triangles[24]. Thus:

$$C_i = \frac{2E_i}{k_i(k_i-1)} \qquad (14)$$

In MBVG, the first neighbors of the node i are the ones seen by that node, i.e. the visibility degree($k_i$) of that node. In order to analysing behavior of systems using clustering coefficient of MBVG, we focus on degrees is more than and equal to 5 i.e. $k \geq 5$. First we calculate it for low degrees and then generalize them for all MBVGs. For a clear analysis, we propose the Tables 1 corresponding to $x_0 = 1$ observer bits. The table 1 is in accord with two intermediate and side states of the observer bit.

According to the Tables 1, it is not hard to detect a precise order for the number of configurations and edges as well as for the clustering coefficient and the probability distribution of the configuration($\beta(k)$). The $\beta(k)$ is equal to $\frac{2}{k-1}$ in the side states and is equal to $\frac{k-3}{k-1}$ in the intermediate states. Also, the number of edges in these states are equal to $2k-5$ for side states and $2k-6$ for intermediate states. Consequently, according to the Eq. 15, we have, $C_S = \frac{4k-10}{k(k-1)}$ for side states and $C_M = \frac{4k-12}{k(k-1)}$ for intermediate states. With regard to the above contents, the visibility degree of nodes ($k$) in the MBVG can be defined as following:

$$k = \begin{cases} \frac{F_{\pm}(C_S)+C_S}{2C_S} & \text{Side states} \\ \frac{F_{\pm}(C_M)+C_M}{2C_M} & \text{Intermediate states} \end{cases} \qquad (15)$$

where $F_{\pm}(C_S)$ and $F_{\pm}(C_M)$ are equal to $4 \pm (C_S^2 - 32C_S + 16)^{\frac{1}{2}}$ and $4 \pm (C_M^2 - 40C_M + 16)^{\frac{1}{2}}$, respectively.

With regard to the obtained results, we are going to calculate the clustering coefficient probability P(C) in different conditions. We must carry out the calculations for two situations: intermediate and side states. In general, when $k \geq 5$, in order to compute the clustering coefficient probability P(C), it would be sufficient to multiply the visibility degree probability distribution P(k) by the one for the configuration $\beta(k)$. As mentioned earlier, we are going to consider first side and second the intermediate states as follows:

### 3.2.1 Side states

The states that observer bit located at side a bounding bit (see table 1). With regard to Eqs. 13 and 16, we have,

$$P_S(k) = \beta_S(k) \times P(k \geq 5) = 2g(k)\, p_1^3 (1-p_1)^{k-2} = P\left(\frac{F_\pm(C_S) + C_S}{2C_S}\right).$$

Assuming the following conditions:

$$\begin{cases} k_+ = \frac{F_+(C_S)+C_S}{2C_S} & k_+ \geq 5 \\ k_- = \frac{F_-(C_S)+C_S}{2C_S} & k_- \geq 5 \end{cases}$$

and, $k_+, k_- \in Z$.
we can get the subsequent relation for clustering coefficient probability:

$$P(C_S) = 2g\left(\frac{F_+(C_S)+C_S}{2C_S}\right) p_1^3 (1-p_1)^{\frac{F_+(C_S)-3C_S}{2C_S}} \tag{16}$$

where $0 < C_A \leq 0.5$.

### 3.2.2 Intermediate states

The states that observer bit located at between the inner bits (see table 1). With regard to Eqs. 13 and 16, we have,

$$P_M(k) = \beta_M(k) \times P(k \geq 5) = (k-3)g(k)\, p_1^3 (1-p_1)^{k-2} = P\left(\frac{F_\pm(C_M)+C_M}{2C_M}\right).$$

Assuming the following conditions:

$$\begin{cases} k_+ = \frac{F_+(C_M)+C_M}{2C_M} & k_+ \geq 5 \\ k_- = \frac{F_-(C_M)+C_M}{2C_M} & k_- \geq 5 \end{cases}$$

and, $k_+, k_- \in Z$.
If these conditions are true,

$$P(C_M) = \sum_{F(C_M)=F_-(C_M)}^{F_+(C_M)} \left(\left(\frac{F(C_M)-5C_M}{2C_M}\right) g\left(\frac{F(C_M)+C_M}{2C_M}\right) p_1^3 (1-p_1)^{\frac{F(C_M)-3C_M}{2C_M}}\right) \tag{17}$$

where $0 < C_M \leq 0.4$ and and if any of the above mentioned conditions($k_+, k_- \in Z$ and $k_+, k_- \geq 5$) are violated, the probability will be zero.

Numerical results shows that the probability distribution of clustering coefficient $P(C)$ of the MBVG corresponding to time series of above examples (the examples of pervious section (3.1)) can be predicted by Eqs. 16 and 17 (see Fig. 6 and Fig 7).

### 3.3 Probability distribution of the visibility length and the mean of the visibility length

The visibility length is defined for the "one" observer. It is the number of intermediate zero bits between two consecutive one-observers in a way that the two one-observers are visible to each other. In Graph terminology, it corresponds to the order of the MPG graph which is a subgraph of MBVG. According to the Eq. 5 and the above contents, we have,

$$P_{1\dashrightarrow 1}(n) = g(n) \times T_{P_{MBS}}(n)$$

where $g(n)$ and $T_{P_{MBS}}(n)$ are the probability density function and the total of possible configurations probability from visibility of $n$ (that is function from multiplying elements of transition probability matrix $P_{MBS}$), respectively. Consequently, it be defined as follows:

$$T_{P_{MBS}}(n) = P(\overbrace{00...00}^{n}) = \overbrace{P_{0\to 0} \times ... \times P_{0\to 0}}^{n}$$

then,

$$P_{1\dashrightarrow 1}(n) = g(n) \times (1-p_1)^n. \tag{18}$$

The value of $g(n)$ for the systems is different and can be classified as following:

**Uncorrelated systems :** The $g(n)$ is constant value. For $n = 1$, $g(n) = 1$ and for $n \geq 2$ is as follows:

$$g(n) = \begin{cases} \alpha > 1 & p_1 > 0.5 \\ \alpha = 1 & p_1 = 0.5 \\ \alpha < 1 & p_1 < 0.5 \end{cases}$$

**Correlated systems :** For correlated stochastic systems,

$$g(n) = \begin{cases} 1 & n = 1, 2 \\ \alpha\, e^{-(n\,\sigma^2)^d} & n \geq 3 \end{cases}$$

and also, for chaotic systems,

$$g(n) = \begin{cases} 1 & n = 1 \\ \alpha\, e^{-(n\,\sigma^2)^d} & n \geq 2 \end{cases}$$

where $\alpha$ and $\sigma^2$ are constant value and variance of the markov-binary sequence, respectively. The value of $d$ is proportional with correlation exponent $\tau$ in correlation stochastic systems and correlation dimension $D$ in chaotic systems.

The probability distribution of the visibility length equals the probability distributions for the MPG. Numerical results confirms that the probability distribution for the MPGs of the MBVG corresponding to time series of above examples (the examples of pervious section (3.1)) can be predicted by Eq. 18 (see Fig. 8 and Fig 9).
According to the above results, the mean visibility length for uncorrelated systems can be defined as follows:

$$L(N) \approx \sum_{n=1}^{N} n P_{1\dashrightarrow 1}(n) = g(n) \sum_{n=1}^{N} n(1-p_1)^n.$$

In which, N is the number of bits in an markov-binary sequence. This can be reduced to:

$$L(N) = g(n) \frac{(p_1 - 1)\left((1-p_1)^N (1 + N p_1) - 1\right)}{p_1^2}. \tag{19}$$

where the $g(n)$ is constant value. This equation gives the mean graph order of $MPG_n$ that equal the mean visibility length for uncorrelated systems. Numerical results confirms that the mean graph order of $MPG_n$ of the MBVG corresponding to time series of uncorrelated systems can be predicted by Eq. 19 (see Fig. 10).

## 4 Network analysis of the human heartbeat dynamics

Recently, the human heartbeat (RR-interval) dynamics has attracted much attention from researchers. As we know, the RR-interval is the time elapsing between two consecutive R waves in the electrocardiogram. The RR-interval time series are different in the normal and patient subjects (see Fig. 11). Here, we investigate the RR-interval time series of three groups from subjects: Normally, Congestive Heart Failure (CHF) and Atrial Fibrillation (AF)

*14*

subjects. Each RR-interval time series is about 24 hours long (roughly $10^5$ intervals). All of the time series were derived from continuous ambulatory (Holter) electrocardiograms (ECGs) that are available on the PhysioNet web site (http://www.physionet.org/challenge/chaos).

Our analysis are on the fluctuations of time in the RR-interval time series using the MBVA. With regarded to above section, the results shows that the degree distribution $P(k)$ and entropy of degree distribution $H(P(k))$ of the MBVG corresponding to RR-interval time series of normally, CHF and AF subjects can be predicted by Eqs. 12 and 13 (see Figs. 12, 13). Also, using entropy of degree distribution $H(P(k))$, this algorithm can indicate that the RR-interval time series of normally, CHF and AF subjects are uncorrelated, chaotic and correlated stochastic systems, respectively(see Fig. 13).

To better analyze the RR-interval time series, we obtain other properties of the MBVG corresponding to RR-interval time series. results indicate that the probability distribution of clustering coefficient $P(C)$ of the MBVG corresponding to RR-interval time series of normally, CHF and AF subjects can be predicted by Eqs. 16 and 17 (see Fig. 14). Also, the probability distribution of the visibility length $P(n)$ of the MBVG corresponding to RR-interval time series of normally, CHF and AF subjects and the mean of the visibility length $L(N)$ of the MBVG corresponding to RR-interval time series of normally subjects can be predicted by Eqs. 18 and 19 (see Fig. 15 and Fig 16).

## 5 Conclusion

In this work, we introduced markov-binary visibility algorithm to analyze the complex systems. The simple structure of this transformation has caused the more precise results obtained in comparison with other transformations. Our results on topological(statistical) properties of the MBVG confirms this statement. In addition, we illustrated this algorithm analyzing the human heartbeat dynamics. The results indicated that the human heartbeat (RR-interval) time series of normally, Congestive Heart Failure (CHF) and Atrial Fibrillation (AF) subjects are uncorrelated, chaotic and correlated stochastic systems, respectively.

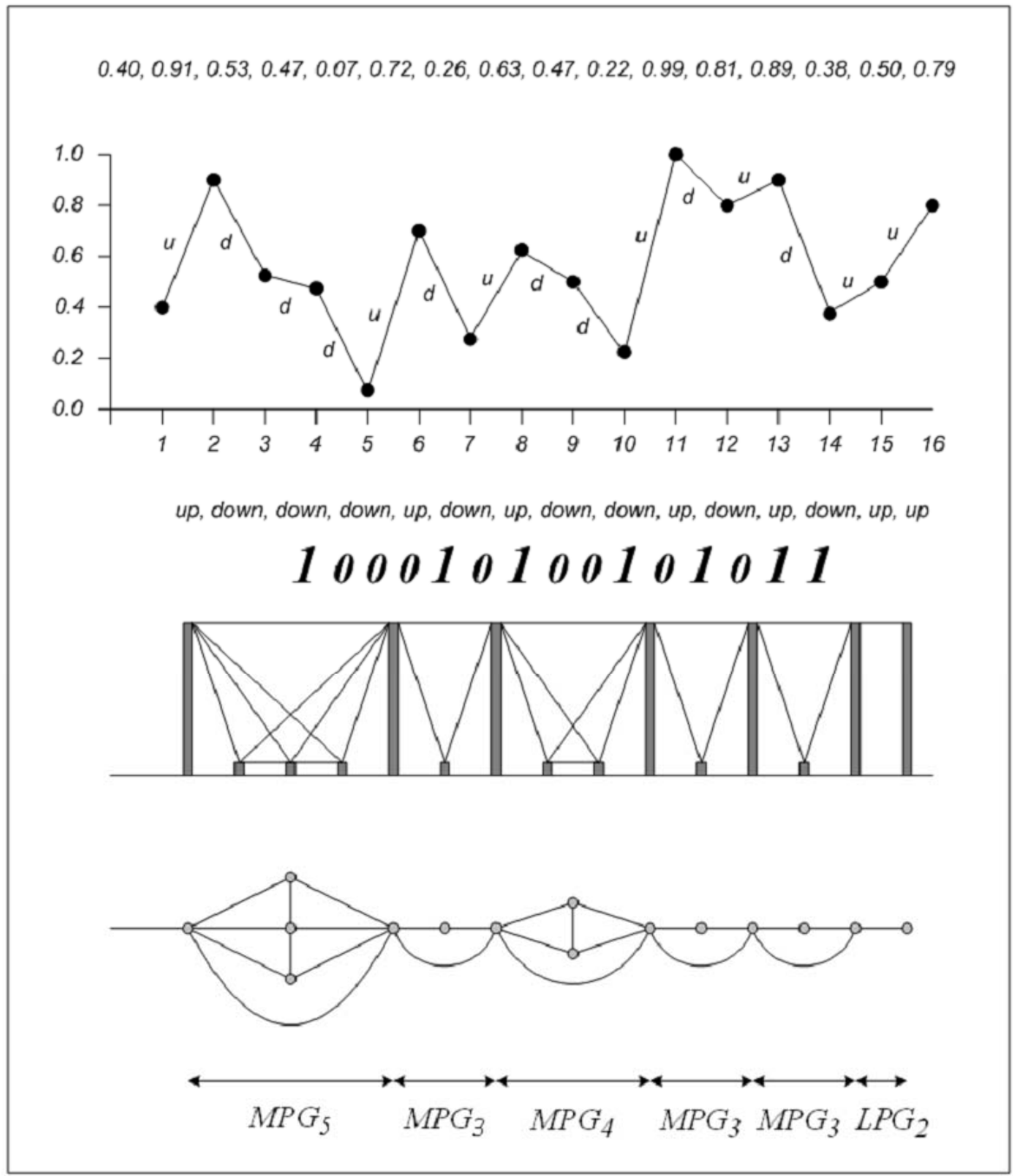

Fig. 1. The steps in converting time series into markov-binary visibility graph. $u$ and $d$ are symbol of states of up and down in markov chain, respectively. $MPG_n$ and $LPG_n$ are subgraphs of the Markov-Binary Visibility Graph(MBVG) that are Maximal Planar Graph and Linear Planar Graph, respectively.

$x_0 = 0$          $x_0 = 1$

$\cdots 0100011011\,1\,0\,1\,0010110011\cdots$   $P_0$   $\cdots 0100011011\,1\,1\,1\,0010110011\cdots$
                $x_{-1}\ x_0\ x_1$                          $x_{-1}\ x_0\ x_1$

For $k = 2$

$x_0 = 0$          $x_0 = 1$

$\cdots 0100011011\,1\,0\,0\,\overset{n}{0\cdots 0}\,1\,0010110011\cdots$   $P_0$   $\cdots 0100011011\,1\,1\,0\,1\,0010110011\cdots$
        $x_{-1}\ x_0\ x_1$      $x_{n+1}$                $x_{-1}\ x_0\ x_1\ x_2$

$\cdots 0100011011\,1\,\overset{n}{0\cdots 0}\,0\,0\,1\,0010110011\cdots$   $P_1$   $\cdots 0100011011\,1\,0\,1\,1\,0010110011\cdots$
        $x_{-(n+1)}$    $x_{-1}\ x_0\ x_1$                 $x_{-2}\ x_{-1}\ x_0\ x_1$

For $k = 3$

$x_0 = 0$          $x_0 = 1$

$\cdots 0100011011\,1\,0\,0\,0\,\overset{n}{0\cdots 0}\,1\,0010110011\cdots$   $P_0$   $\cdots 0100011011\,1\,1\,0\,0\,1\,0010110011\cdots$
     $x_{-2}\ x_{-1}\ x_0\ x_1$     $x_{n+1}$             $x_{-1}\ x_0\ x_1\ x_2\ x_3$

$\cdots 0100011011\,1\,\overset{n_1}{0\cdots 0}\,0\,0\,0\,\overset{n_2}{0\cdots 0}\,1\,0010110011\cdots$   $P_1$   $\cdots 0100011011\,1\,0\,1\,0\,1\,0010110011\cdots$
    $x_{-(n_1+1)}$   $x_{-1}\ x_0\ x_1$   $x_{n_2+1}$          $x_{-2}\ x_{-1}\ x_0\ x_1\ x_2$

$\cdots 0100011011\,1\,\overset{n}{0\cdots 0}\,0\,0\,0\,1\,0010110011\cdots$   $P_2$   $\cdots 0100011011\,1\,0\,0\,1\,1\,0010110011\cdots$
    $x_{-(n+1)}$    $x_{-1}\ x_0\ x_1\ x_2$            $x_{-3}\ x_{-2}\ x_{-1}\ x_0\ x_1$

For $k = 4$

$x_0 = 1$

$P_0$   $\cdots 0100011011\,1\,1\,0\,0\,0\,1\,0010110011\cdots$
                 $x_{-1}\ x_0\ x_1\ x_2\ x_3\ x_4$

$P_1$   $\cdots 0100011011\,1\,0\,1\,0\,0\,1\,0010110011\cdots$
                 $x_{-2}\ x_{-1}\ x_0\ x_1\ x_2\ x_3$

$P_2$   $\cdots 0100011011\,1\,0\,0\,1\,0\,1\,0010110011\cdots$
                 $x_{-3}\ x_{-2}\ x_{-1}\ x_0\ x_1\ x_2$

$P_3$   $\cdots 0100011011\,1\,0\,0\,0\,1\,1\,0010110011\cdots$
                 $x_{-4}\ x_{-3}\ x_{-2}\ x_{-1}\ x_0\ x_1$

For $k = 5$

Fig. 2. The configurations for k=2,3,4,5

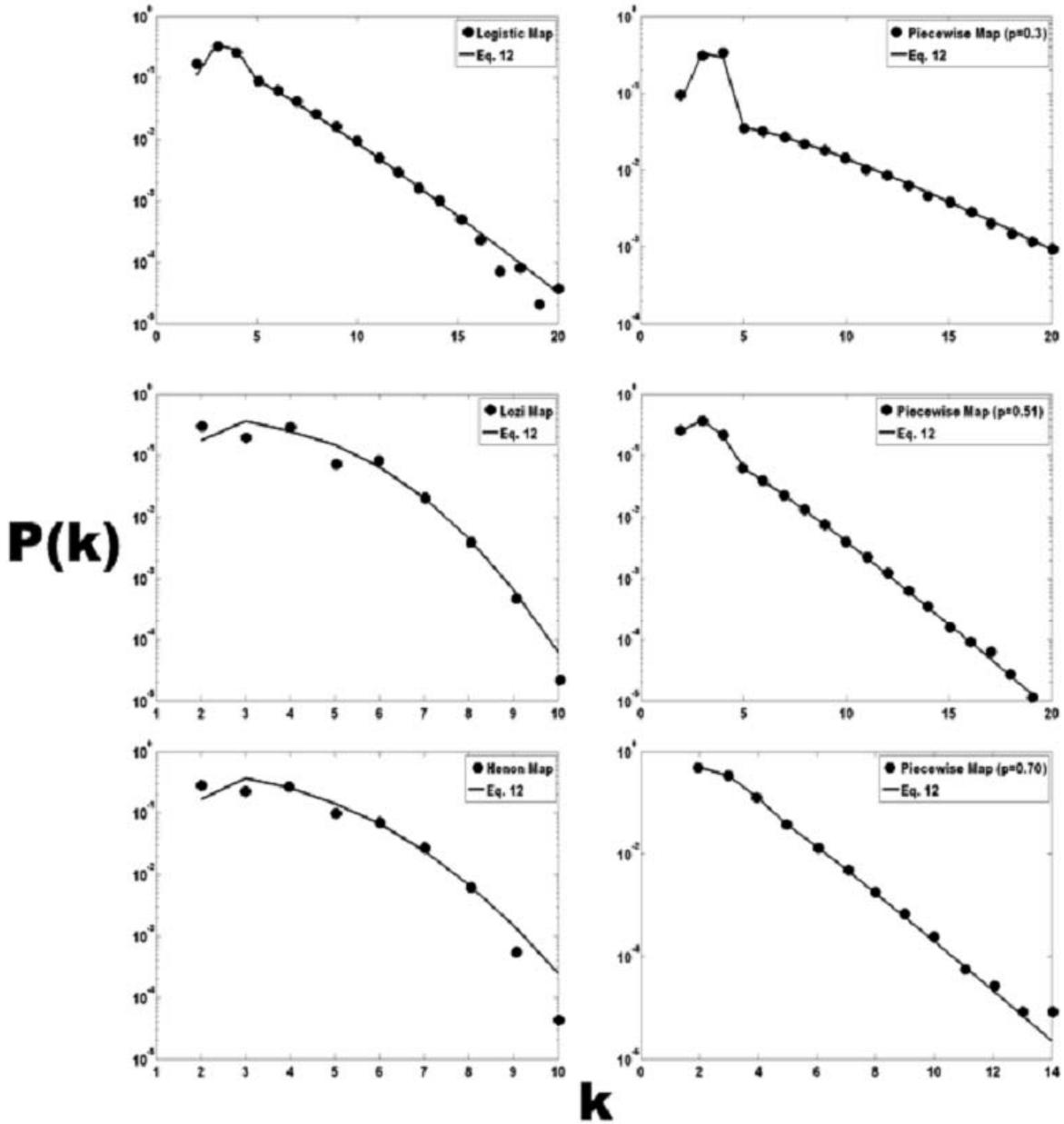

Fig. 3. Exprimental degree distribution of the markov-binary visibility graph corresponding to time series ($10^5 data$) of the chaotic systems (black circles) left column and the uncorrelated systems(piecewise map) (black circles) right column. Solid line corresponding to Eq. 12. In this equation, for chaotic systems $\alpha = 4$ and also, $d = 1D$ for logistic map and $d = 2D$ for henon and lozi maps ($D$ is correlation dimension).

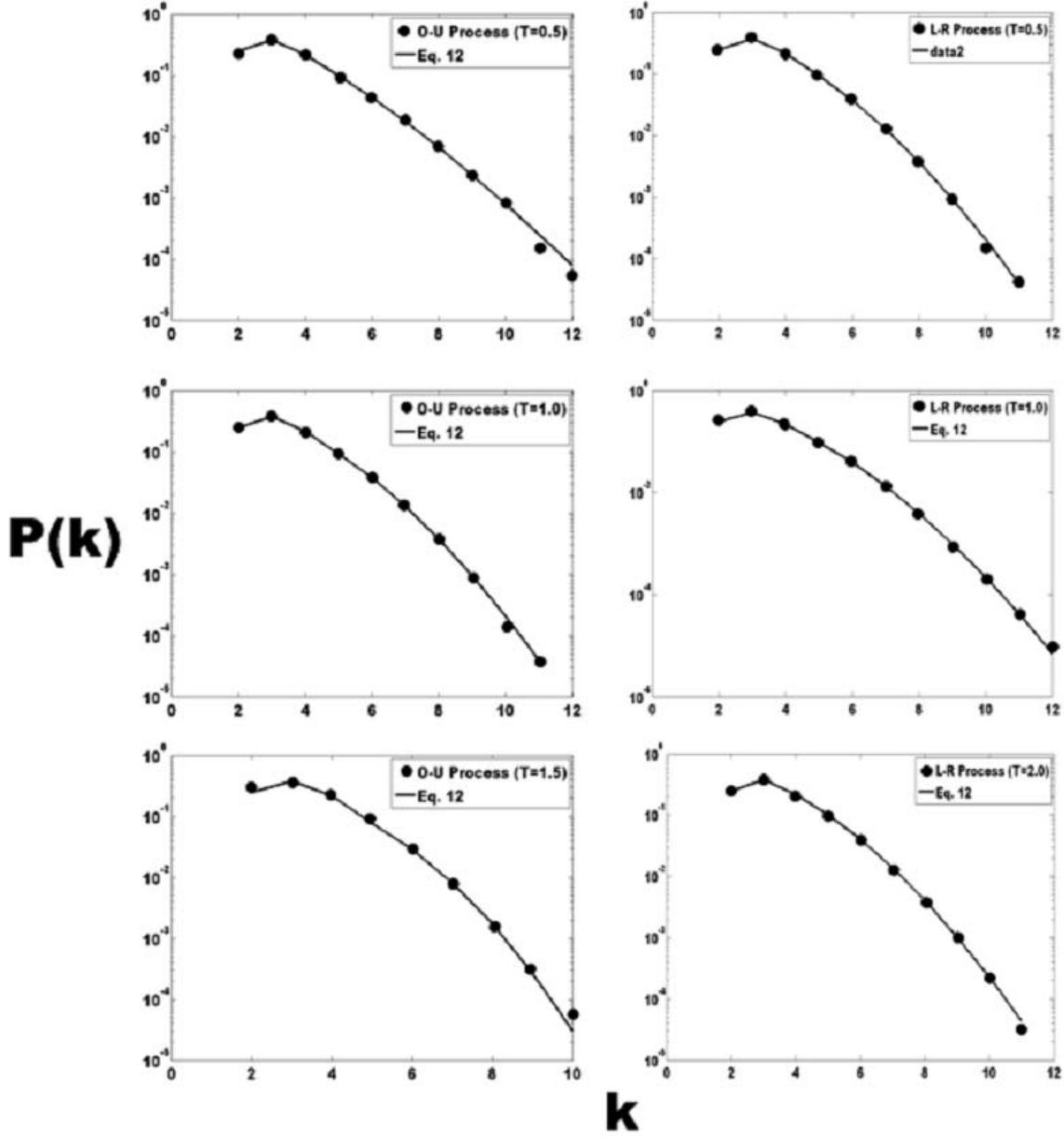

Fig. 4. Exprimental degree distribution of the markov-binary visibility graph corresponding to time series ($10^5 data$) of the correlated stochastic systems (black circles). The Ornstein-Uhlenbeck (O-U) processes are in left column and the Long-Range stochastic (L-R) processes are in right column. Solid line corresponding to Eq. 12. In this equation, for Ornstein-Uhlenbeck (O-U) processes, $d = 1.5$, $\alpha = 3.0$ for $\tau = 0.5$, $d = 2.0$, $\alpha = 2.5$ for $\tau = 1.0$ and $d = 2.5$, $\alpha = 2.0$ for $\tau = 1.5$, for Long-Range stochastic (L-R) processes, $d = 2$ and also, $\alpha = 2.6$ for $\tau = 0.5$, $\alpha = 2.7$ for $\tau = 1.0$ and $\alpha = 2.8$ for $\tau = 2.0$.

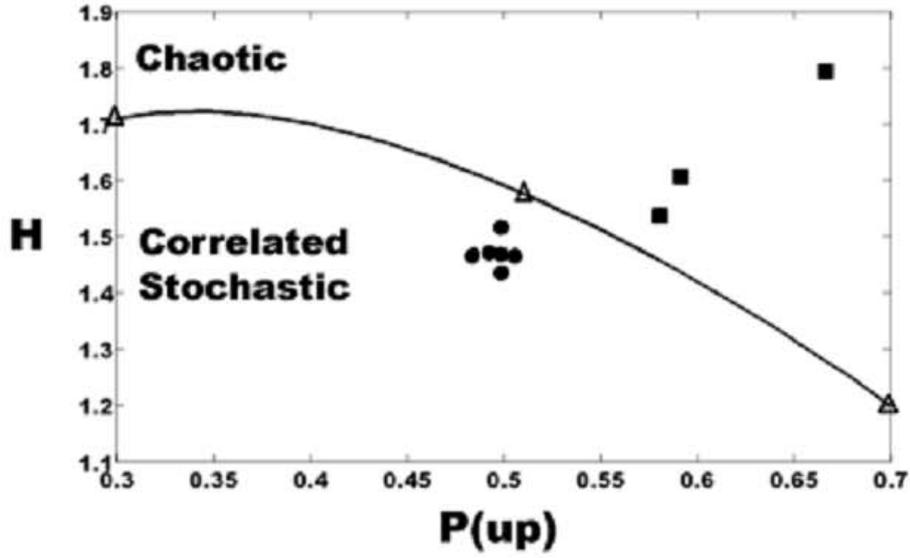

Fig. 5. Exprimental entropy of the degree distribution of the markov-binary visibility graph corresponding to time series ($10^5 data$) of the chaotic systems (black squares), the uncorrelated systems(piecewise map) (triangles) and correlated stochastic systems (black circles). Solid line corresponding to Eq. 13 for uncorrelated systems.

Table 1
Probable configurations for one-observer in intermediate and side states

|  | Side states | Intermediate states |  |  |  | Side states |
|---|---|---|---|---|---|---|
| **E** | 5 | 4 |  | 4 |  | 5 |
| **k=5** | 1<u>1</u>0001 | 10<u>1</u>001 |  | 100<u>1</u>01 |  | 1000<u>1</u>1 |
| **E** | 7 | 6 | 6 | 6 |  | 7 |
| **k=6** | 1<u>1</u>00001 | 10<u>1</u>0001 | 100<u>1</u>001 | 1000<u>1</u>01 |  | 10000<u>1</u>1 |
| **E** | 9 | 8 | 8 | 8 | 8 | 9 |
| **k=7** | 1<u>1</u>000001 | 10<u>1</u>00001 | 100<u>1</u>0001 | 1000<u>1</u>001 | 10000<u>1</u>01 | 100000<u>1</u>1 |
| ⋮ | ⋮ |  | ⋮ |  |  | ⋮ |

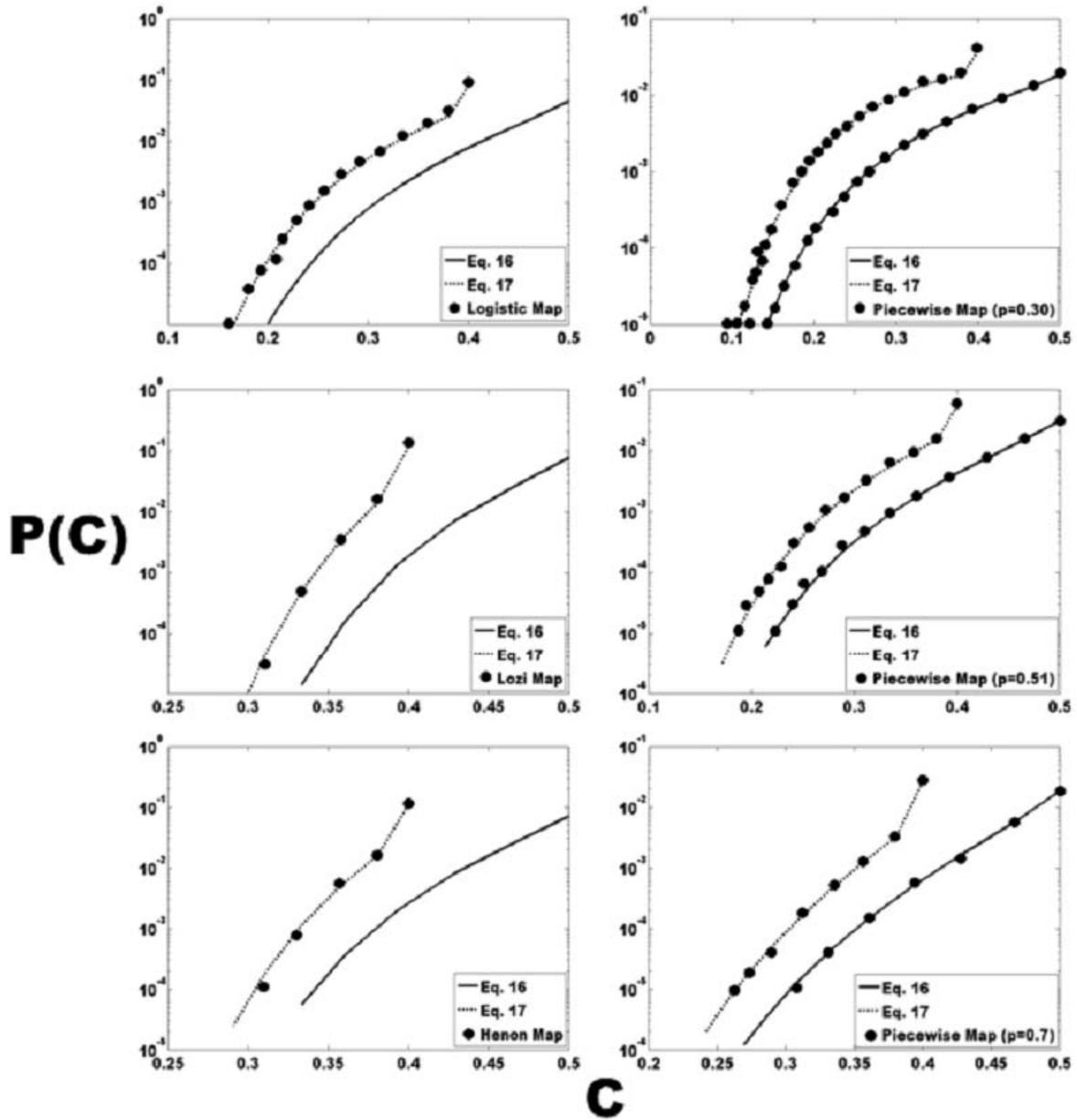

Fig. 6. Exprimental probability distribution of the clustering coefficient of the markov-binary visibility graph corresponding to time series ($10^5 data$) of the chaotic systems (black circles) left column and the uncorrelated systems(piecewise map) (black circles) right column. Solid line corresponding to Eq. 16 and point line corresponding to Eq. 17. In this equations, for chaotic systems $\alpha = 4$ and also, $d = 1D$ for logistic map and $d = 2D$ for henon and lozi maps ($D$ is correlation dimension).

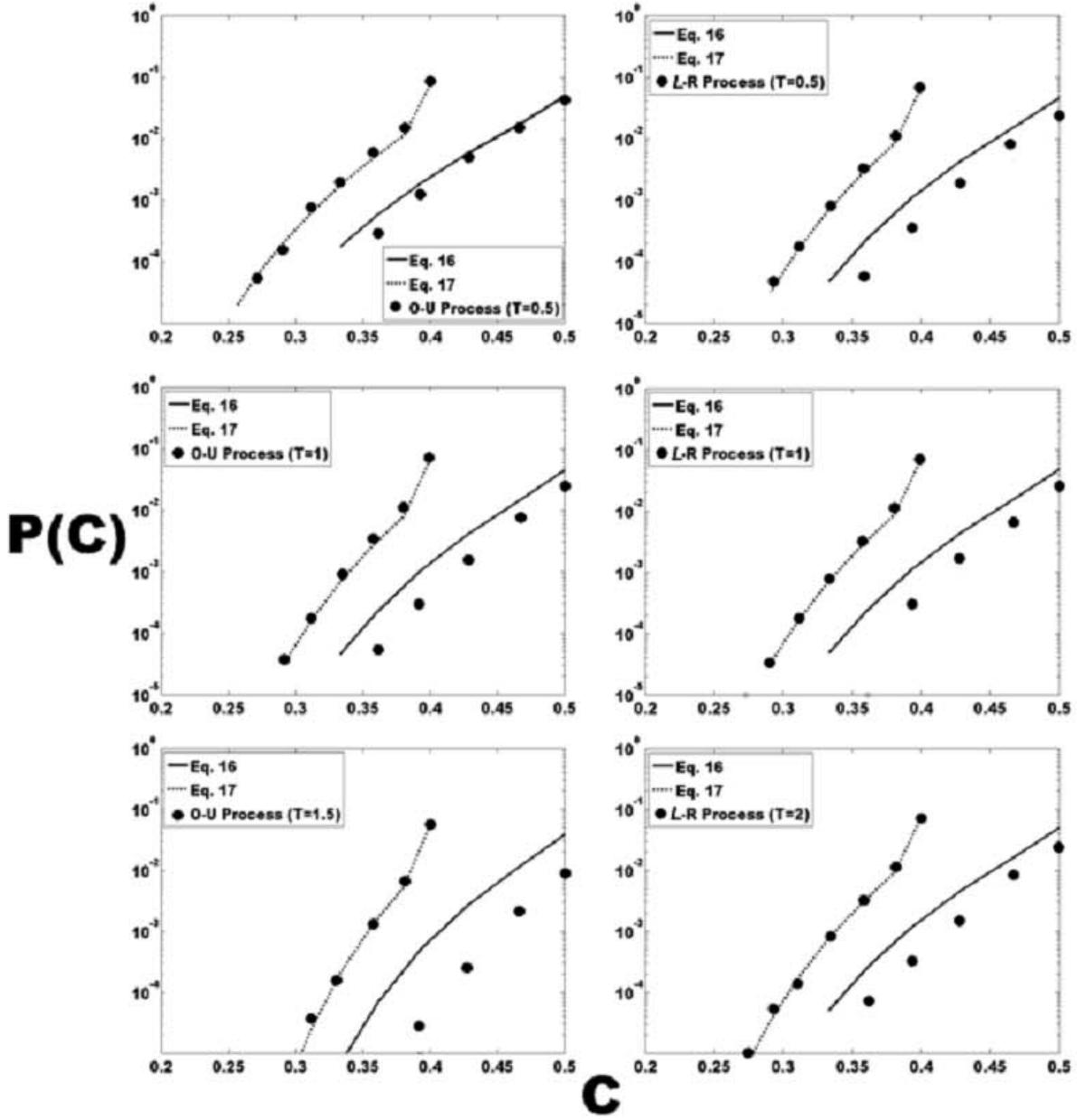

Fig. 7. Exprimental probability distribution of the clustering coefficient of the markov-binary visibility graph corresponding to time series ($10^5 data$) of the correlated stochastic systems (black circles). The Ornstein-Uhlenbeck (O-U) processes are in left column and the Long-Range stochastic (L-R) processes are in right column. Solid line corresponding to Eq. 16 and point line corresponding to Eq. 17. In this equations, for Ornstein-Uhlenbeck (O-U) processes, $d = 1.5$, $\alpha = 3.0$ for $\tau = 0.5$, $d = 2.0$, $\alpha = 2.5$ for $\tau = 1.0$ and $d = 2.5$, $\alpha = 2.0$ for $\tau = 1.5$, for Long-Range stochastic (L-R) processes, $d = 2$ and also, $\alpha = 2.6$ for $\tau = 0.5$, $\alpha = 2.7$ for $\tau = 1.0$ and $\alpha = 2.8$ for $\tau = 2.0$.

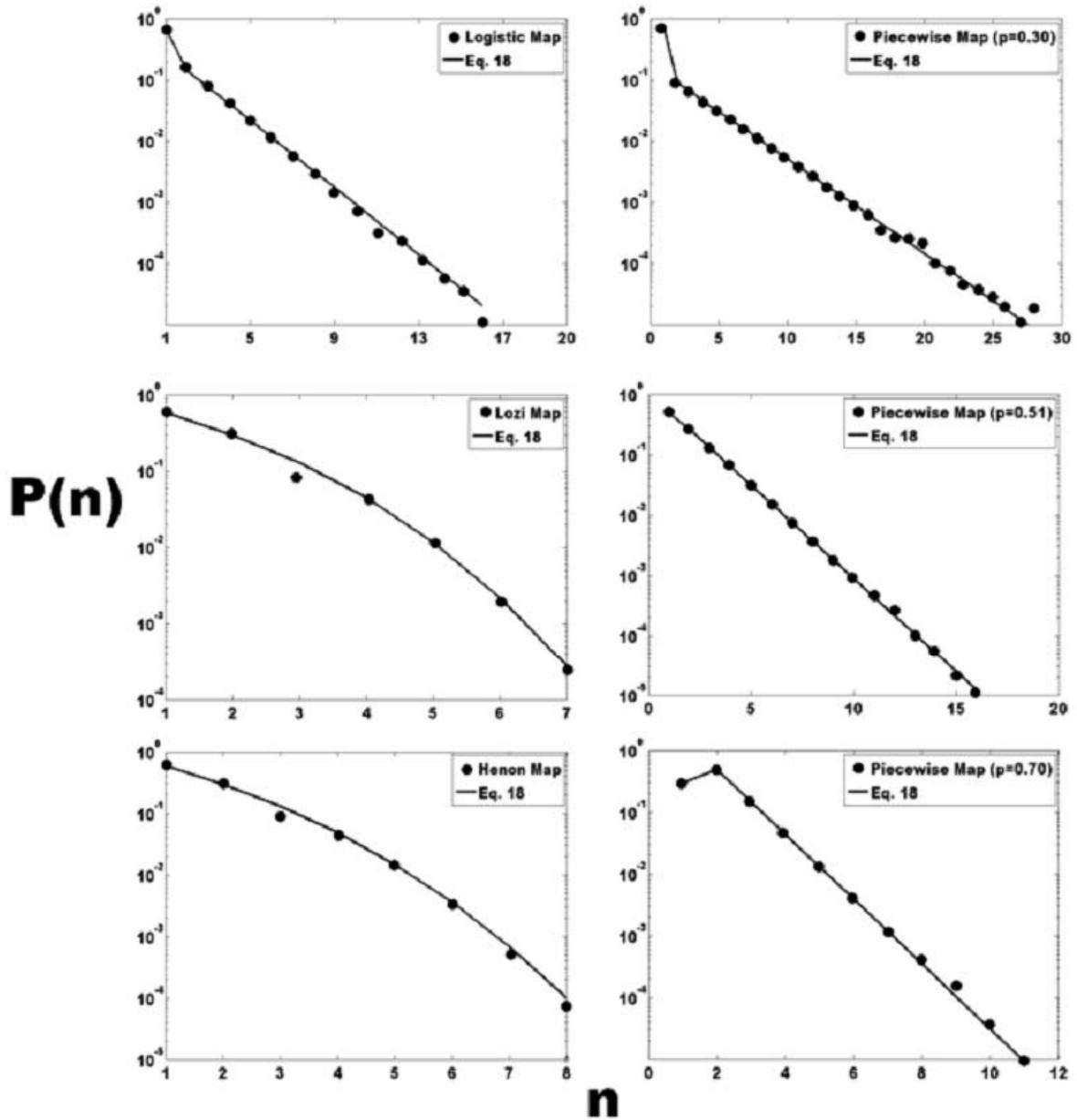

Fig. 8. Exprimental probability distribution for the $MPG_n$ (visibility length) of the markov-binary visibility graph corresponding to time series ($10^5 data$) of the chaotic systems (black circles) left column and the uncorrelated systems(piecewise map) (black circles) right column. Solid line corresponding to Eq. 18. In this equation, $\alpha = 0.18$, $1.10$, $5.50$ for uncorrelated systems(piecewise map) $p = 0.30$, $0.51$, $0.70$, respectively, $\alpha = 1.0$, $d = 1D$ for logistic map and $\alpha = 1.0$, $d = 2D$ for henon and lozi maps ($D$ is correlation dimension).

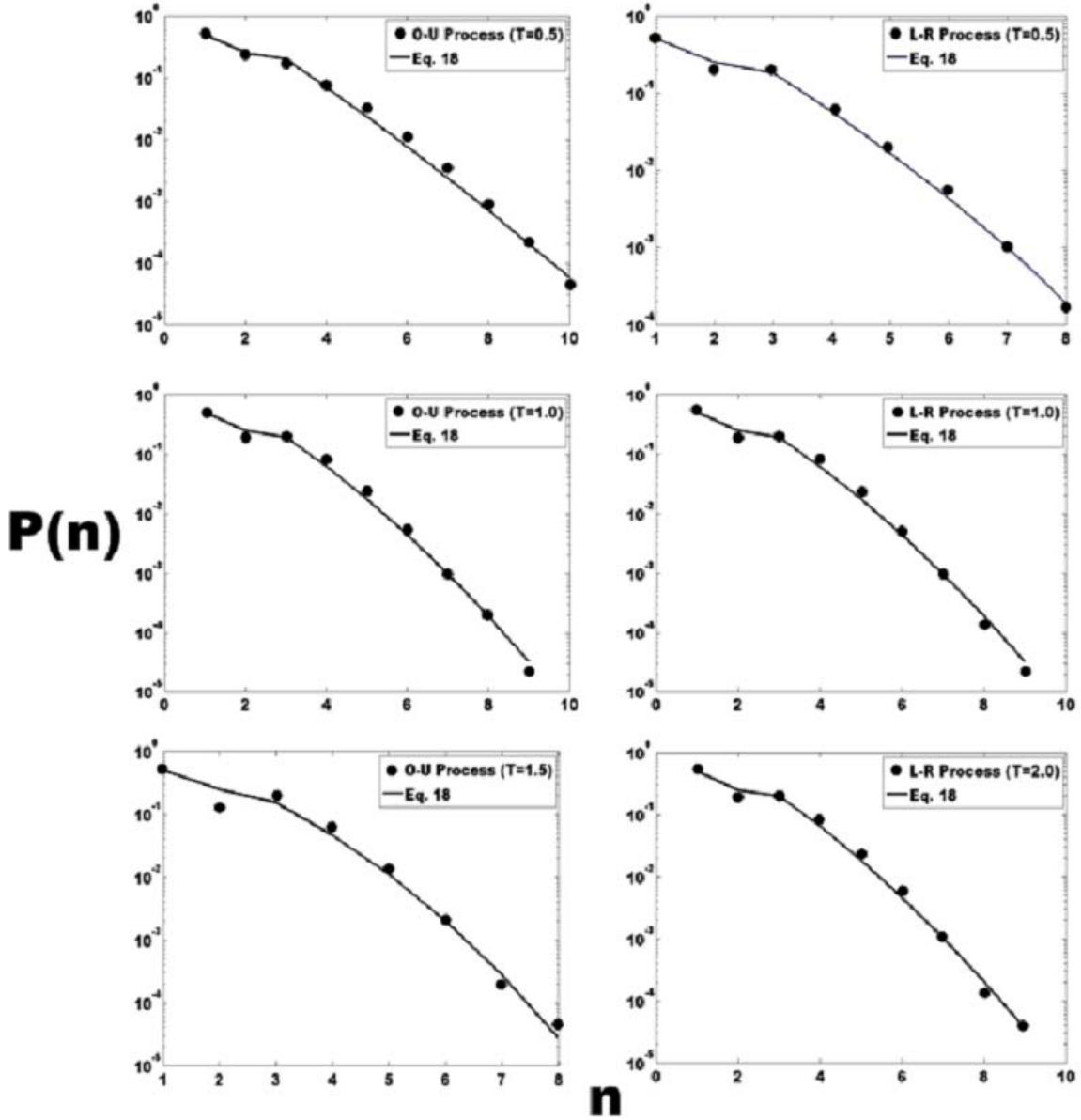

Fig. 9. Exprimental probability distribution for the $MPG_n$ (visibility length) of the markov-binary visibility graph corresponding to time series ($10^5 data$) of the correlated stochastic systems (black circles). The Ornstein-Uhlenbeck (O-U) processes are in left column and the Long-Range stochastic (L-R) processes are in right column. Solid line corresponding to Eq. 18. In this equation, for Ornstein-Uhlenbeck (O-U) processes, $d = 1.5$, $\alpha = 3.0$ for $\tau = 0.5$, $d = 2.0$, $\alpha = 2.5$ for $\tau = 1.0$ and $d = 2.5$, $\alpha = 2.0$ for $\tau = 1.5$, for Long-Range stochastic (L-R) processes, $d = 2$ and also, $\alpha = 2.6$ for $\tau = 0.5$, $\alpha = 2.7$ for $\tau = 1.0$ and $\alpha = 2.8$ for $\tau = 2.0$.

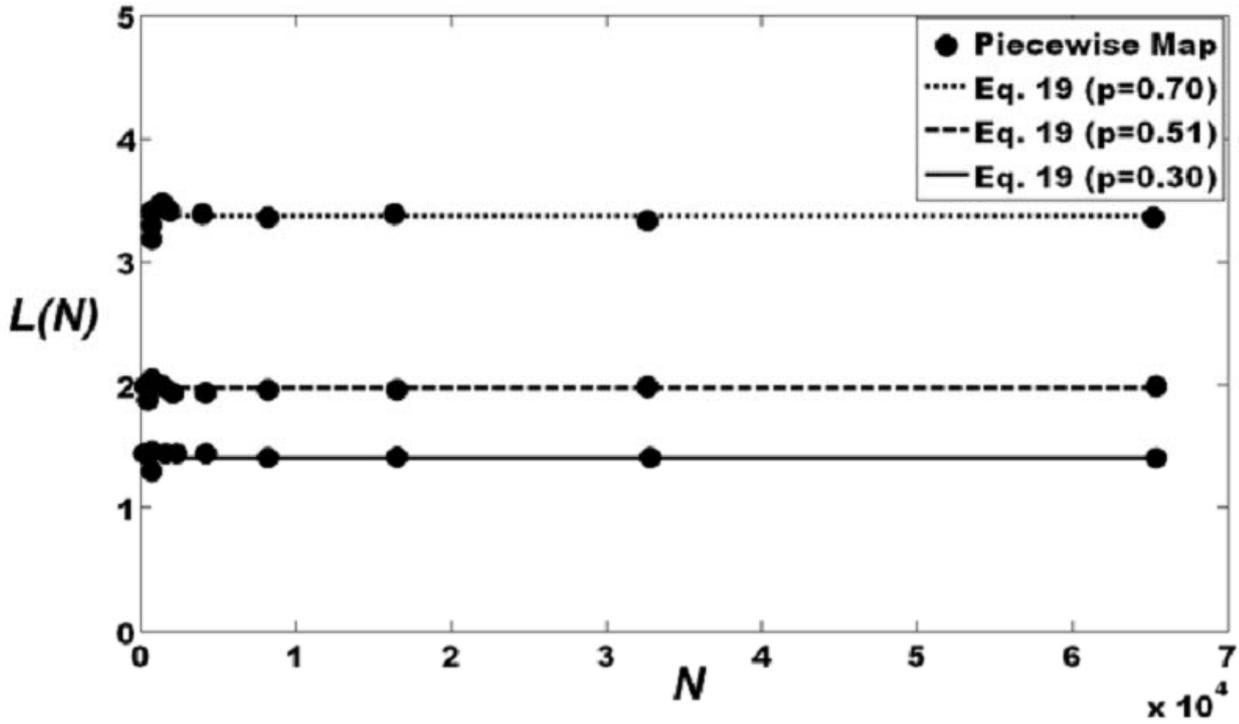

Fig. 10. Exprimental mean subgraphs for the $MPG_n$ (mean of the visibility length) of the markov-binary visibility graph corresponding to time series ($N = 2^7, ..., 2^{16} data$) of the uncorrelated systems(piecewise map) (black circles). The lines corresponding to Eq. 19. In this equation, $\alpha = 0.18, 1.10, 5.50$ for uncorrelated systems(piecewise map) $p = 0.30, 0.51, 0.70$, respectively.

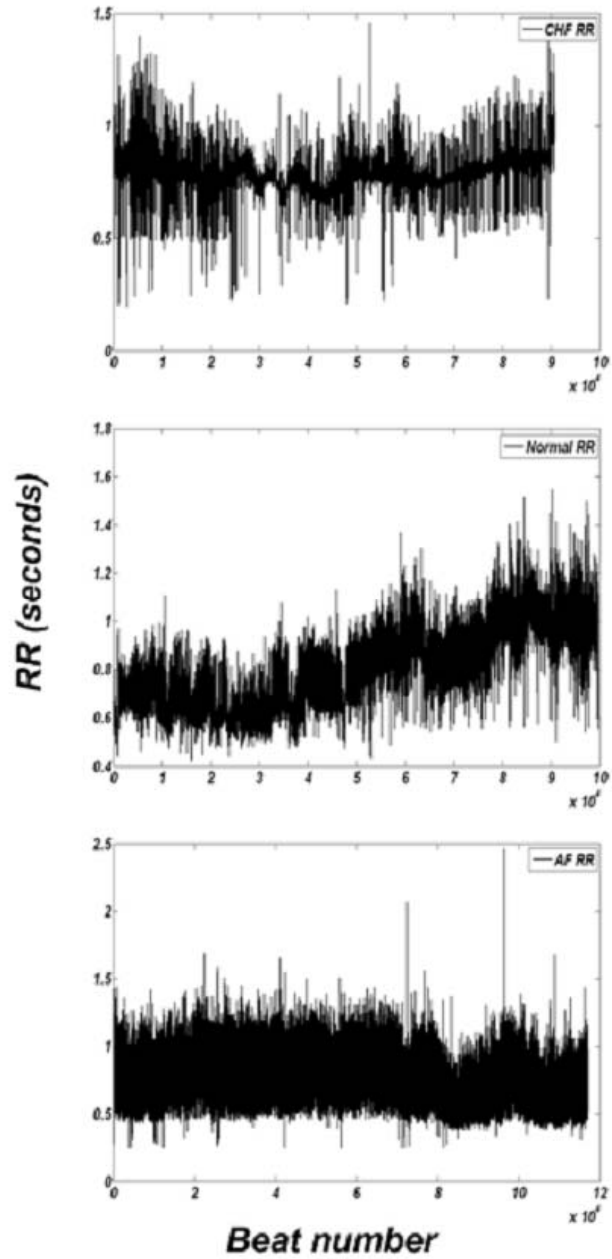

Fig. 11. Three RR-interval time series of three groups from subjects: Normally, Congestive Heart Failure (CHF) and Atrial Fibrillation (AF) subjects.

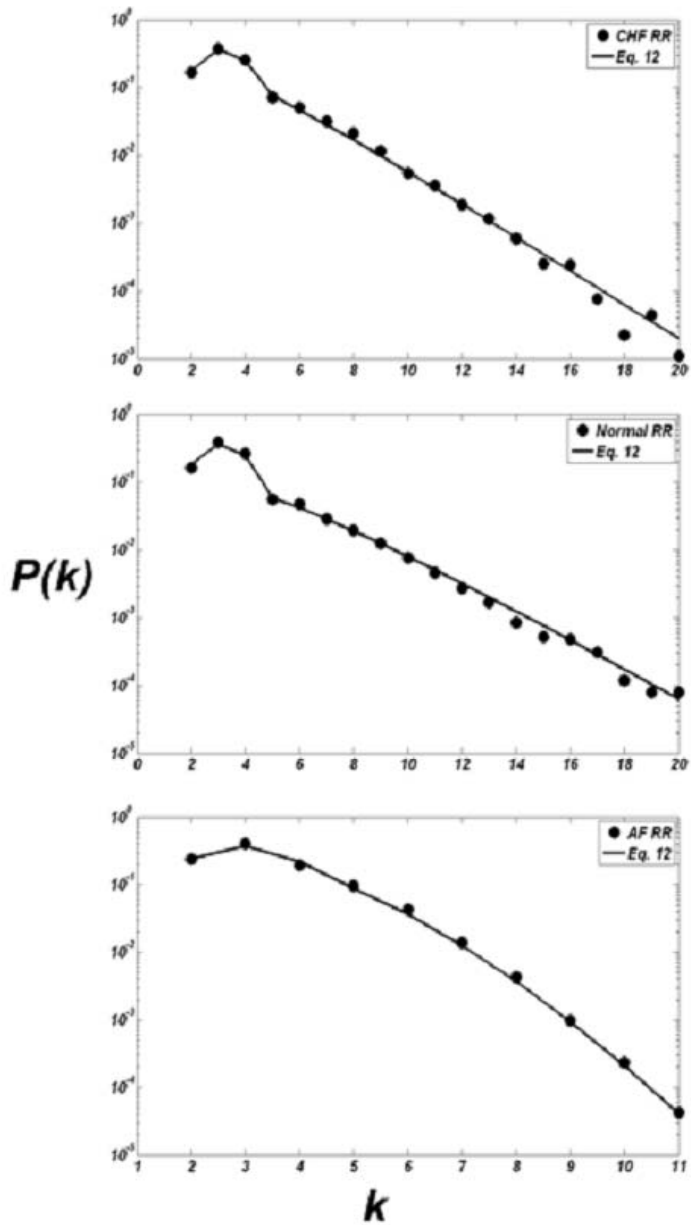

Fig. 12. Exprimental degree distribution of the markov-binary visibility graph corresponding to RR-interval time series ($10^5 data$) of three groups from subjects: Normally, Congestive Heart Failure (CHF) and Atrial Fibrillation (AF) subjects (black circles). Solid line corresponding to Eq. 12. In this equation, $\alpha = 3$ and $d = 0.6$ for RR-interval time series of CHF subjects, also, $\alpha = 2.5$ and $d = 2$ for RR-interval time series of AF subjects.

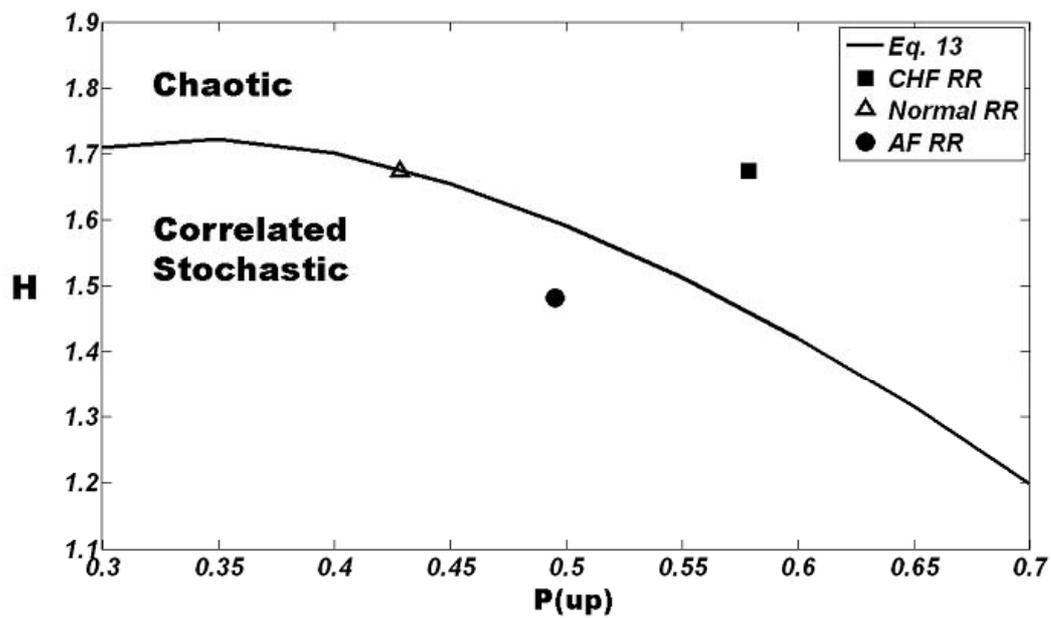

Fig. 13. Exprimental entropy of the degree distribution of the markov-binary visibility graph corresponding to RR-interval time series ($10^5 data$) of three groups from subjects: Normally (triangle), Congestive Heart Failure (CHF) (black square) and Atrial Fibrillation (AF) subjects (black circle). Solid line corresponding to Eq. 13 for uncorrelated systems.

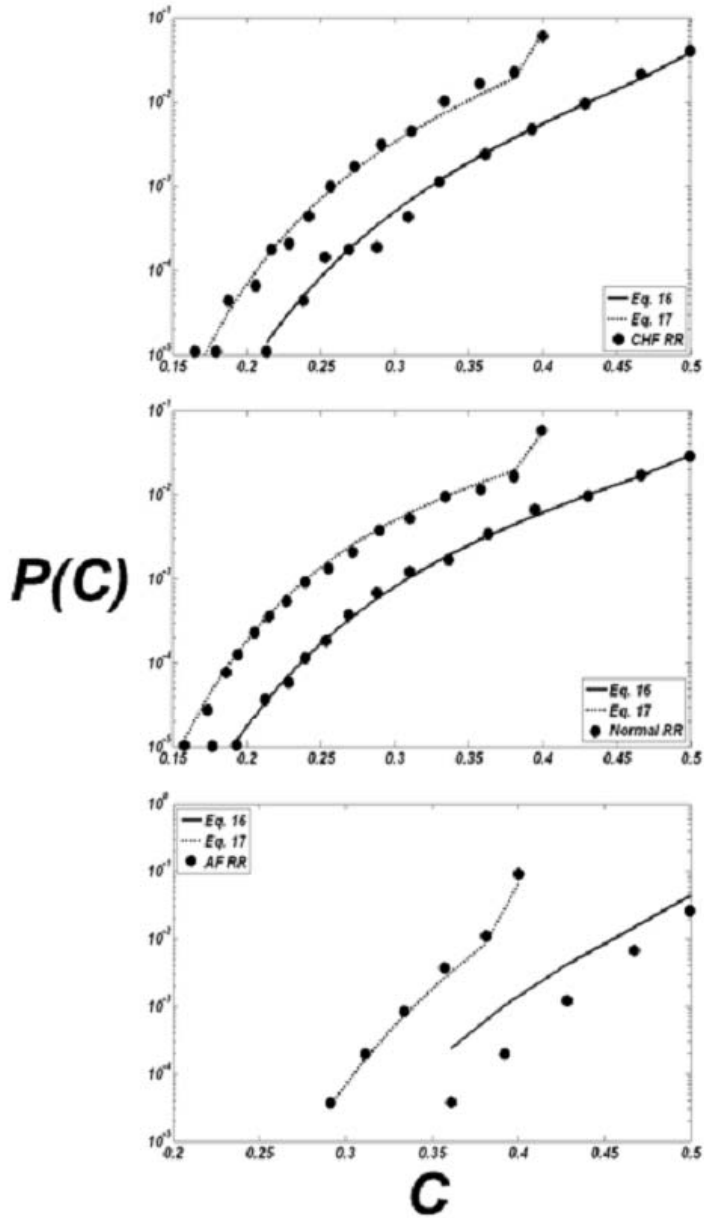

Fig. 14. Exprimental probability distribution of the clustering coefficient of the markov-binary visibility graph corresponding to RR-interval time series ($10^5 data$) of three groups from subjects: Normally, Congestive Heart Failure (CHF) and Atrial Fibrillation (AF) subjects (black circles). Solid line corresponding to Eq. 16 and point line corresponding to Eq. 17. In this equation, $\alpha = 3$ and $d = 0.6$ for RR-interval time series of CHF subjects, also, $\alpha = 2.5$ and $d = 2$ for RR-interval time series of AF subjects.

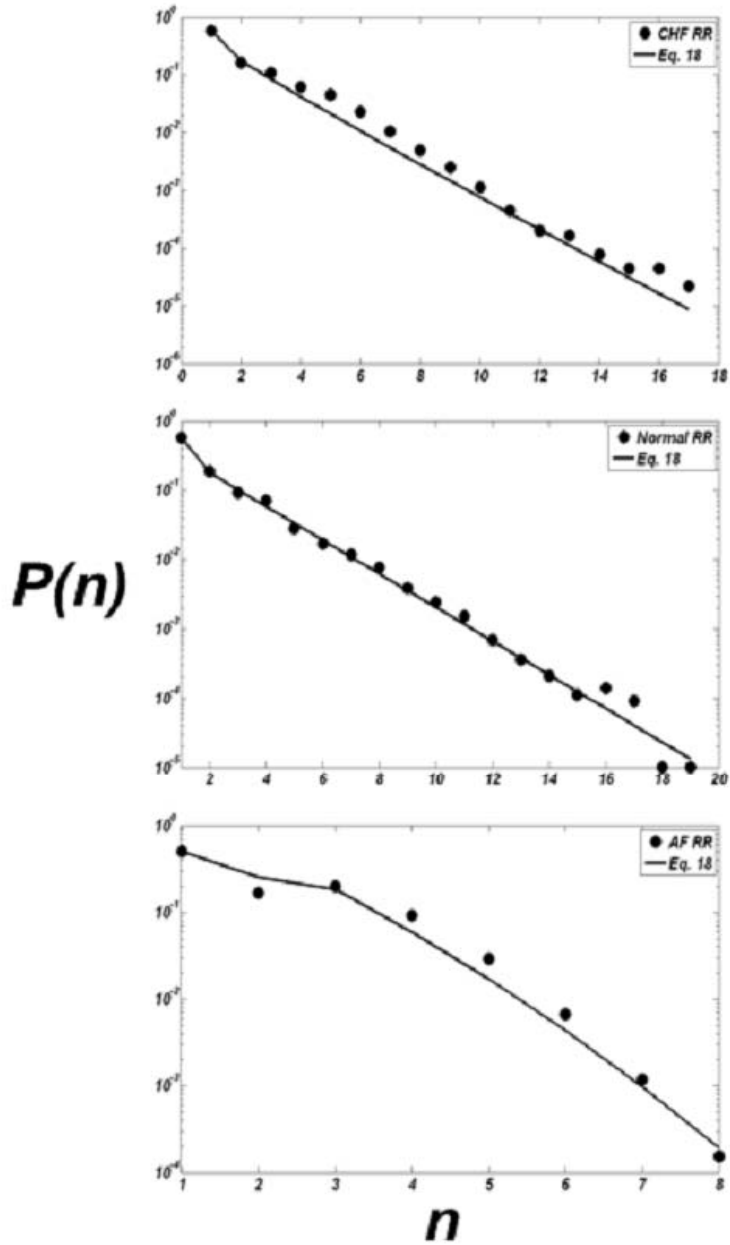

Fig. 15. Exprimental probability distribution for the $MPG_n$ (visibility length) of the markov-binary visibility graph corresponding to RR-interval time series ($10^5 data$) of three groups from subjects: Normally, Congestive Heart Failure (CHF) and Atrial Fibrillation (AF) subjects (black circles). Solid line corresponding to Eq. 18. In this equation, $\alpha = 0.55$ for normally subjects, $\alpha = 1.0$, $d = 0.6$ for RR-interval time series of CHF subjects and $\alpha = 2.5$, $d = 2$ for RR-interval time series of AF subjects.

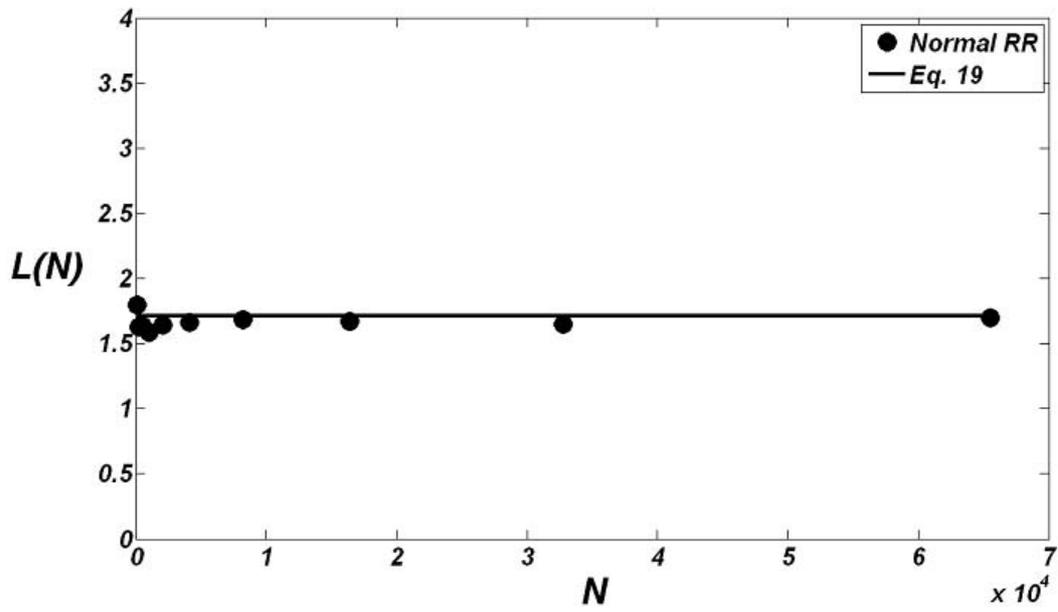

Fig. 16. Exprimental mean subgraphs for the $MPG_n$ (mean of the visibility length) of the markov-binary visibility graph corresponding to RR-interval time series ($N = 2^7, ..., 2^{16} data$) of normally subjects (black circles). The lines corresponding to Eq. 19. In this equation, $\alpha = 0.55$ for normally subjects.